# Accurate 3D comparison of complex topography with terrestrial laser scanner : application to the Rangitikei canyon (N-Z)

Dimitri Lague[1,2], Nicolas Brodu[3], Jérôme Leroux[1]

1: Géosciences Rennes, Université Rennes 1, CNRS, Campus de Beaulieu, 35042 Rennes, France. Email: Dimitri.Lague@univ-rennes1.fr, tel: +33 2 2323 56 53, fax: +33 2 2323

2: Dpt of Geological Sciences, University of Canterbury, Christchurch, New-Zealand.

3: Institut de Physique de Rennes, Université Rennes 1, CNRS, Campus de Beaulieu, 35042 Rennes, France.

## Abstract

Surveying techniques such as Terrestrial Laser Scanner have recently been used to measure surface changes via 3D point cloud (PC) comparison. Two types of approaches have been pursued: 3D tracking of homologous parts of the surface to compute a displacement field, and distance calculation between two point clouds when homologous parts cannot be defined. This study deals with the second approach, typical of natural surfaces altered by erosion, sedimentation or vegetation between surveys. Current comparison methods are based on a closest point distance or require at least one of the PC to be meshed with severe limitations when surfaces present roughness elements at all scales. To solve these issues, we introduce a new algorithm performing a direct comparison of point clouds in 3D. The method has two steps: (1) surface normal estimation and orientation in 3D at a scale consistent with the local surface roughness ; (2) measurement of the mean surface change along the normal direction with explicit calculation of a local confidence interval. Comparison with existing methods demonstrates the higher accuracy of our approach, as well as an easier workflow due to the absence of surface meshing or DEM generation. Application of the method in a rapidly eroding meandering bedrock river (Rangitikei river canyon) illustrates its ability to handle 3D differences in complex situations (flat and vertical surfaces on the same scene), to reduce uncertainty related to point cloud roughness by local averaging and to generate 3D maps of uncertainty levels. We also demonstrate that for high precision survey scanners, the total error budget on change detection is dominated by the point clouds registration error and the surface roughness. Combined with mm-range local georeferencing of the point clouds, levels of detection down to 6 mm (defined at 95 % confidence) can be routinely attained in situ over ranges of 50 m. We provide evidence for the self-affine behaviour of different surfaces. We show how this impacts the calculation of normal vectors and demonstrate the scaling behaviour of the level of change detection. The algorithm has been implemented in an open source software freely available. It operates in complex full 3D case and can also be used as a simpler and more robust alternative to DEM differencing for the 2D cases.

*Revised version, 01 February 2013*

# 1. Introduction

Terrestrial laser scanner (TLS) and photogrammetric techniques are increasingly used to track the evolution of natural surfaces in 3D at an unprecedented resolution and precision. Existing applications encompass landslide and rockfall dynamics (Wawrzyniec et al., 2007; Teza et al., 2008; Abellán et al., 2009, 2010), coastal cliff erosion (e.g. Rosser et al., 2005; Olsen et al., 2011), braided rivers evolution (e.g. Milan et al., 2007), river bank erosion (e.g. O'Neal and Pizzuto, 2011) or debris flows impacts (Schürch et al., 2011). Unravelling surface change in these contexts requires the comparison of two point clouds in 2D or 3D. Two type of measurements exists : (i) computation of a displacement field based on the identification of corresponding elements within successive point clouds (Teza et al., 2007; Monserrat and Crosetto, 2008; Aryal et al., 2012); (ii) distance measurement between two clouds as used in change detection and volumetric calculation problems (e.g. Girardeau-Montaut et al., 2005; Rosser et al., 2005). This latter calculation does not assume correspondence between point cloud elements and measures a net surface erosion or sedimentation central to many problems in geomorphology (e.g. Milan et al., 2007; Schürch et al., 2011). This work address the 3D change detection problem in the context of rough complex topographies lacking corresponding elements among successive point clouds.

Compared to the high level of maturity reached by instruments, solutions to perform point cloud comparison in 3D are scarce and hardly adapted to complex natural surfaces. For instance, fig. 1a shows a TLS survey of a meandering incised river (Rangitikei, N-Z) which is a 3D manifold surface exhibiting horizontal (e.g. river bed) and vertical surfaces (e.g. cliff, block faces). A regular 2D grid representation of this surface (as in a Digital Elevation Model) would necessarily introduce bias on vertical or overhanging parts, but also limit the resolution of fine scale details due to the fixed grid size. Similarly a difference of point clouds along the vertical direction would biased measurements of horizontal bank retreat or cliff erosion, highlighting the need for a computation of distances along a direction locally normal to the surface. The Rangitikei example is also characterized by surfaces of various roughnesses from flat rock faces to rough gravel beds and rockfall deposits (fig. 1b). High roughness surfaces generate occlusion patterns (i.e. missing data) that depend on the viewpoint and introduce complexity in the point cloud comparison (Girardeau-Montaut et al., 2005; Zeibak and Filin, 2007; Hodge, 2010). It also makes the calculation and orientation of surface normals dependent on the scale at which it is performed (Mitra and Nguyen, 2003; Lalonde et al., 2005). Finally, a confidence interval should be attached to each distance measurement and used to test if a statistically significant change has occurred. This confidence interval should factor in the cumulative effects of point cloud alignment uncertainty, roughness effects and errors related to the instrument measurement (Gordon et al., 2001; Abellán et al., 2009; Wheaton et al., 2009; Schürch et al., 2011). To date, comparing complex topographies such as the Rangitikei river in 3D with an explicit calculation of spatially variable confidence intervals is not feasible.

This paper seeks to fill this gap and presents a new method called Multiscale Model to Model Cloud Comparison (M3C2) which combines for the first time three key characteristics :

- it operates directly on point clouds without meshing or gridding.
- it computes the local distance between two point clouds along the normal surface direction which tracks 3D variations in surface orientation.
- it estimates for each distance measurement a confidence interval depending on point cloud roughness and registration error.

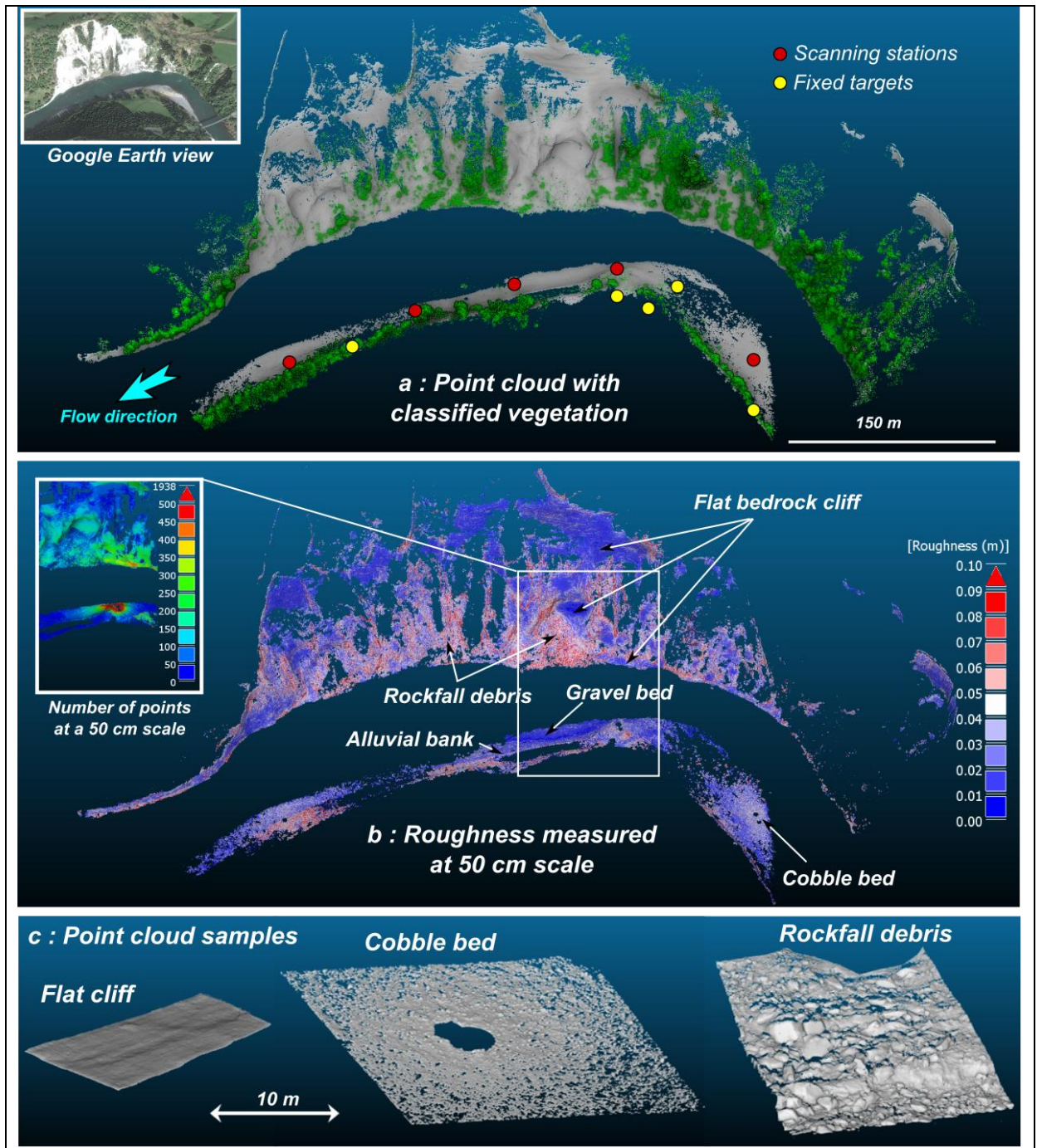

**Figure 1: A**: Scan data and aerial view of a meander of the Rangitikei river in New-Zealand (coordinates: -39.827681,175.783842). Apart from the vegetation which is suppressed here with the CANUPO technique (Brodu and Lague, 2012), the surface consists of an alluvial river bed with grainq ranging from sand to meter size boulders, rock cliff faces and rockfall debris whose typical geometry is shown in c. **B:** Surface roughness without vegetation measured at a scale of 0.5 m (see section 3.1 for the calculation method) with indications of the various excerpts used in the geometrical analysis of the statistical properties of the surfaces in section 5.1. **C:** View of 3 of the sample surfaces illustrating the very large variability in surface roughness. These three samples are used in the analysis of the impact of roughness on the various calculations of the M3C2 algorithm in section 5. Visualisations with Cloudcompare (EDF R&D, 2011).

The paper is divided into 5 sections. We first review existing methods of point cloud comparison and the source of uncertainties. In the second section we present the M3C2 algorithm and a comparison of its performance against existing techniques on synthetic point clouds. The third section presents the data acquisition and registration methods used for the test data. We use 2 surveys of a 500 m long meander reach of the Rangitikei river gorge in New-Zealand (fig. 1) exhibiting a wide range of surface roughness characteristics and surface change. In the fourth section we analyze the roughness properties of typical surfaces of the test data and its impact on the calculation output. It is used to discuss the choice of the two calculation parameters that must be defined by the user. A typical output of the M3C2 algorithm on the test data is shown. We demonstrate how precise registration techniques combined with the new algorithm yield levels of surface change detection down to 6 mm in the best cases. Finally we discuss the limitations and possible improvements of the method.

## 2. Existing distance measurement methods and sources of uncertainties

In the following we present the main advantages and drawbacks of the 3 main approaches used to measure the distance between two point clouds in the context of geomorphologic applications. We then discuss the main source of uncertainties relevant to the point cloud comparison problem.

### 2.1 DEM of difference (DoD)

DEM of difference is the most common method of point cloud comparison in earth sciences when the large scale geometry of the scene is planar (i.e. river bed, cliff). The two point clouds are gridded to generate DEMs either directly if the large scale surface is near horizontal (e.g. channel bed: Lane et al., 2003; Milan et al., 2007; Wheaton et al., 2009; Schürch et al., 2011) or after rotation (e.g. cliff, river banks : Rosser et al., 2005; Abellán et al., 2010; O'Neal and Pizzuto, 2011). The two DEMs are then differentiated on a pixel-by-pixel basis which amounts at measuring a vertical distance . This technique is very fast and now includes explicit calculation of uncertainties related to point cloud registration, data quality and point cloud roughness (e.g. Brasington et al., 2000; Lane et al., 2003; Wheaton et al., 2009; Schürch et al., 2011), Yet the DoD technique suffers from two major drawbacks:

- it cannot operate properly on 3D environments such as the Rangitikei river (fig. 1a) as a DEM cannot cope with overhanging parts (cliffs and bank failures, large blocks) and decreases information density proportionally to surface steepness (i.e. vertical surfaces cannot be described by a DEM). While in figure 1a, a DoD could be computed for the near horizontal gravel bed, the resolution of surface change calculation on near vertical parts of the cliff would be limited by the DEM resolution. In that case, surface change should be measured along the normal direction of the surface which is nearly horizontal.
- even if the surface is 2D at large scale, gridding TLS point cloud data is a difficult task for rough surfaces (Hodge, 2010; Schürch et al., 2011). A rough surface is always 3D at some scale and the corresponding point cloud acquired from the ground will have missing data due to occlusion. Missing data will be interpolated which introduces an uncertainty on the grid elevation. Because point density and roughness can be extremely variable, the choice of a representative elevation in a cell is not simple (Schürch et al., 2011; Rychkov et al., 2012). Finally, the fixed DEM resolution imposes a limit on the level of detail retained from the raw data which can be a strong limitation on surfaces exhibiting very different characteristic scales (i.e. wide range of grain sizes).

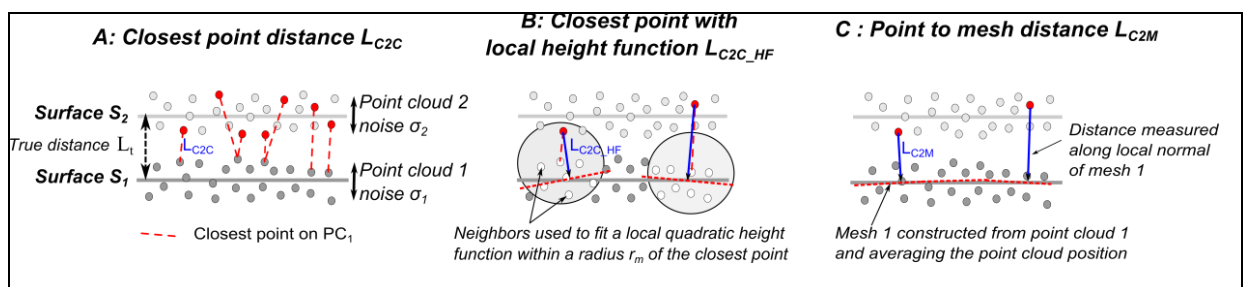

**Figure 2**: **Existing 3D comparison methods between two point clouds PC$_1$ and PC$_2$**. Each point cloud is characterized by a roughness ($\sigma_1$ and $\sigma_2$ that are not *a priori* identical) that is a combination of instrumented related noise and surface roughness. In this example we assume that S$_1$ is the reference surface that has been displaced along the local normal of S$_1$ by a true distance $L_t$. A distance is calculated for each point of S$_2$. **A: Simplest cloud-to-cloud distance $L_{C2C}$**, based on the closest point distance. For small $L_t$, $L_{C2C}$ is dependent on the roughness and point density of PC$_1$ and PC$_2$. **B: Closest point distance to local model distance $L_{C2C\_HF}$** (C2C_HF). A local height function (e.g. a plane) is computed using the neighbouring points within a radius $r_m$ of the closest point in PC$_1$. This provides a better approximation of the true position of $S_1$ but does not entirely remove the sensitivity to outliers and point density in PC$_1$ owing to the choice of the closest point in PC$_1$. The measurement is also dependent on $\sigma_2$. **C: Cloud-to-mesh (C2M) distance $L_{C2M}$.** PC$_1$ is meshed and the distance between each point of PC$_2$ and S$_1$ is computed along the local normal of S$_1$. If the mesh correctly approximates the average position of S$_1$, then $L_{C2M} = L_t \pm \sigma_2$.

Because the DOD technique remains inherently limited to 2D surfaces, we do not test it subsequently as we are only interested in 3D methods. We note that the M3C2 algorithm can be used in 2D by imposing a vertical normal calculation which is equivalent to the DoD method without the need for gridding.

## 2.2 Direct cloud-to-cloud comparison with closest point technique (C2C)

This method is the simplest and fastest direct 3D comparison method of point clouds as it does not require gridding or meshing of the data, nor calculation of surface normals (Girardeau-Montaut et al., 2005). For each point of the second point cloud, a closest point can be defined in the first point cloud. In its simplest version, the surface change is estimated as the distance between the two points (C2C, fig. 2A). Improvements can be obtained by a local model of the reference surface either by an height function (C2C_HF, fig. 2B) or by a least square fit of the closest point neighbours (Girardeau-Montaut et al., 2005). This technique is also used in cloud matching techniques such as the ICP (Besl and McKay, 1992; Yang and Medioni, 1992). Application of this technique in the context of the Rangitikei river overcomes the limitation of the DoD technique with respect to 3D features such as overhanging parts. Yet, as shown in fig. 2A and 2B, the measured distance is sensitive to the clouds roughness, outliers and point spacing. This sensitivity is further studied in section 3.3. For this reason, the technique was developed for rapid change detection on very dense point clouds rather than accurate distance measurement (Girardeau-Montaut et al., 2005). An evolution of this technique in which the normal orientation is used to sign the difference is available in a commercial software (Polyworks from Innovmetrics). No spatially variable confidence interval is currently calculated with this technique.

## 2.3 Cloud-to-mesh distance or cloud-to-model distance (C2M)

This approach is the most common technique in inspection software. Surface change is calculated by the distance between a point cloud and a reference 3D mesh or theoretical model (Cignoni and Rocchini, 1998), see also Monserrat and Crosetto (2008) and Olsen et al. (2010) for recent reviews). This approach works well on flat surfaces as a mesh corresponding to the average reference point cloud position can be constructed (fig. 2C) (e.g. Kazhdan et al., 2006). However, creating a surface mesh is complex for point clouds with significant roughness at all scales or missing data due to occlusion. It generally requires time-consuming manual inspection. As for the DoD technique, interpolation over missing data introduces uncertainties that are difficult to quantify. Mesh construction also smooth out some details that may be important to assess local roughness properties. Spatially variable confidence intervals and test for statistically significant change where discussed by Van Gosliga et al. (2006) in the case of tunnel deformation (cloud to model comparison) but have not been addressed in the case of rough natural surfaces with the cloud to mesh techniques.

## 2.4 Sources of uncertainty in point cloud comparison

Identifying the sources of uncertainty in point cloud comparison is essential to construct confidence intervals. Three main sources can be identified :

1. **Position uncertainty of point clouds:** the latest generation of time of flight scanners now offers a range precision (i.e. range noise) down to 1.4 mm, a range accuracy below 1 mm and positioning uncertainty of 4 mm at 50 m (Boehler et al., 2003; Mechelke et al., 2007; Bae et al., 2009). These characteristics increase with distance and incidence angle (e.g. Soudarissanane et al., 2009, 2011) and may depend on surface characteristics for some instruments (e.g. Boehler et al., 2003). No simple model can currently account for the position uncertainty of a point within a point cloud (Bae et al., 2009; Soudarissanane et al., 2011), but we note that the range noise can be estimated locally at any distance by measuring the cloud roughness provided that the scanned surface is planar. Because this noise is normally distributed, averaging $n$ samples will reduce the standard error on the mean point position by a factor $\sqrt{n}$ (Abellán et al., 2009).

2. **Registration uncertainty between the point clouds**: except for the rare case of a scanning instrument replaced exactly in the same position, the coordinate systems of the two clouds have a systematic error that is a complex function of the method used to georeference the two clouds, the number of stations to generate a single survey and the scanning instrument characteristics (Lichti et al., 2005; Salvi et al., 2007; Bae and Lichti, 2008; Olsen et al., 2009, 2011). Two techniques are currently used to register point clouds: ground controls points (GCP) that are fixed between surveys or resurveyed with an independent method (e.g. theodolite, GPS) (e.g. Alba et al., 2006; Milan et al., 2007) and cloud matching techniques that use overlapping parts of the cloud (Besl and McKay, 1992; Yang and Medioni, 1992; Salvi et al., 2007; Bae and Lichti, 2008; Olsen et al., 2011; Schürch et al., 2011). Registration quality can be assessed with independent control points or reference surfaces not used in the registration. The typical registration errors obtained in natural environments of scale similar to the Rangitikei river scene (~ 500 m) are of the order of a few cm : 3 cm (Rosser et al., 2005), 7.9 cm (Olsen et al., 2009), 2-3 cm (Schürch et al., 2011), 2 cm (Milan et al., 2007), 1.7 cm (Abellán et al., 2010). Alba *et al.*, (2006) reports errors down to 5.7 mm using fixed GCPs. The registration error can be anisotropic (Bae and Lichti, 2008) and may not be spatially uniform if the distribution of registration constraints is not homogeneous in 3D. The instrument accuracy and precision play a critical role in the final registration error.

3. **Surface roughness related errors:** these are caused by the difficulty to reoccupy exactly the same scanning position between surveys, by the occlusion due to roughness and by the positioning uncertainties inherent to TLS (Hodge et al., 2009a; Hodge, 2010; Schürch et al., 2011). These effects cause the spatial sampling of rough surfaces to never be identical between surveys (e.g. Hodge et al, 2009a). Hence, even if the surface did not change, a small difference will systematically be erroneously measured. A correctly defined confidence interval should then discard the difference as non-statistically significant. When a true change of the surface occurs between surveys, the occlusion pattern might change and induces artificially large surface changes in places which are "suddenly" visible from the scanner position (e.g. Girardeau-Montaut et al., 2005; Zeibak and Filin, 2007). Sharp features will also generate spurious points called mixed-points that will create local outliers on the surface (e.g. Boehler et al., 2003; Hodge et al., 2009a). These can be partially removed by point cloud preprocessing (Hodge et al., 2009a; Brodu and Lague, 2012). Finally, roughness will also affect the calculation of the surface normal orientation (Mitra and Nguyen, 2003; Lalonde et al., 2005; Bae et al., 2009) which might lead to a potential overestimation of the distance between the two clouds. The contribution of roughness effects to the total error budget is the least well constrained of all.

## 3. Methodology and synthetic tests

We first describe how the M3C2 operates to measure the distance between two points clouds, how confidence intervals are estimated and compare its calculations to existing techniques using synthetic point clouds. Assuming that the two clouds correspond to successive surveys, the first one will be called the *reference* cloud, and the second one the *compared* cloud.

## 3.1 Presentation of the M3C2 algorithm

### 3.1.1 Calculation on core points

We use a set of calculation "core" points for which one distance and confidence interval is calculated. The core points will generally be a sub-sampled version of the reference cloud (e.g., by setting a minimum point spacing), but all calculations use the original raw data. The notion of "core" points was introduced in Brodu and Lague (2012) to significantly speed up the calculations. It accounts for the fact that calculation results are generally needed at a lower, more uniform spatial resolution (e.g. 10 cm) than the raw irregular point spacing of high density scans (e.g. 1 cm or less). Core points can also be viewed as a kind of Region Of Interest analysis in which point cloud comparison will be performed : they can have any spatial organization such as a regular grid whose calculation results can be easily transformed into a raster format. The point clouds themselves may also be used directly (each point is a core point) if so desired.

### 3.1.2 Step 1 : Calculation of surface normals in 3D

For any given core point *i*, a normal vector is defined for each cloud by fitting a plane to the neighbours $NN_i$ of that cloud that are within a radius $D/2$ of *i* (fig. 3a). Each normal is oriented positively towards the closest of a set of user-defined "orientation" points (e.g. generally the various scanning positions). The standard deviation of the distance of the neighbours $NN_i$ to the best fit plane is recorded and used as a measure of the cloud roughness $\sigma_i(D)$ at scale $D$ in the vicinity of *i*. It is also known as the detrended roughness (e.g. Heritage and Milan, 2009; Rychkov et al., 2012). We discuss in section 3.3.1 how the normal scale $D$ can be defined according to the local roughness of the cloud. The algorithm offers the option to compute surface normals on a sub-sampled version of the point clouds to speed-up the calculation while retaining a good accuracy.

The algorithm offers the possibility to use either the normal estimated on the reference cloud, on the compared cloud, or the average of both normal direction. When the two clouds are locally co-planar this does not change the measured distance. When a change in surface orientation has occurred the measured distance will depend on the chosen reference normal (fig 3b). Using the average of the two normal mitigates this effect and makes the measurement reversible. The choice of the normal is ultimately imposed by the user depending on its application. In geomorphology and in the absence of correspondence between the two clouds, the distance will generally be calculated using the normal estimated on the reference point cloud. This is justified by the fact that geomorphic processes tend to depend on surface geometry (e.g. topographic slope for river processes, hillslope erosion, rock stability). Hence, owing to causality the measured change is more likely resulting from the initial geometry, rather than the final one.

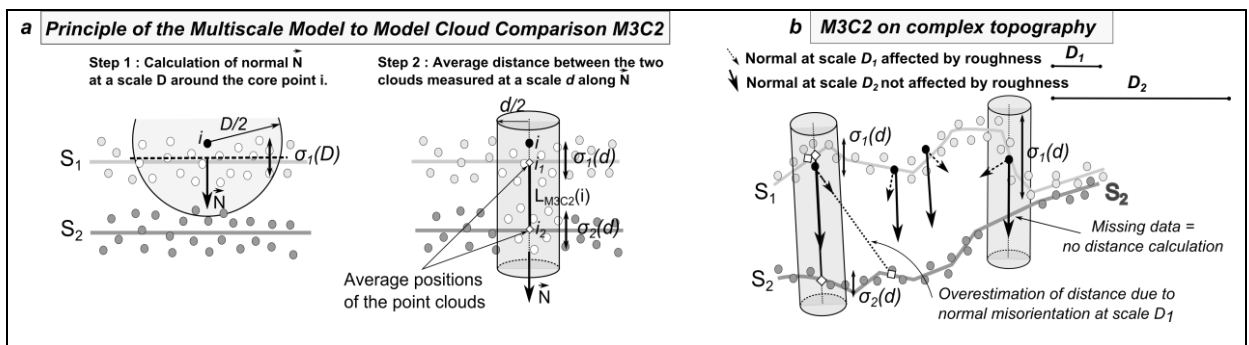

**Figure 3**: Description of the M3C2 algorithm and the two user-defined parameters *D* (normal scale) and *d* (projection scale). a : simple configuration illustrating the two steps of the algorithm : Step 1 : In this example the normal is estimated from cloud 1 (see the main text for possible alternatives and a symmetric distance definition if so desired). In that latter case the scale at which the cloud is the most planar will be selected. Step 2 : 2 sub-clouds are defined by the intersection of the reference and compared clouds with a cylinder of diameter *d* and axis (*i*, $\vec{N}$ ). Each sub-cloud is projected on the cylinder axis which gives a distribution of distances along the normal direction. These are used to define the mean (or median) position of each cloud $i_1$ and $i_2$. The local point cloud roughness $\sigma_1(d)$ and $\sigma_2(d)$ and size $n_1$ and $n_2$ of the 2 sub-clouds are subsequently used to estimate a parametric local confidence interval. b : application on a complex topography : for rough surfaces, if the normal is measured at a scale too small with respect to the surface roughness characteristics ($D_1$), its orientation will strongly varies. This will tend to overestimate the distance between the two clouds. In this example a larger normal scale $D_2$ yields a more uniform normal orientation. The subsequent distance calculation is more representative of the average orthogonal distance between cloud 1 and cloud 2. This figure also illustrates a case in which some data is missing in the second surface yielding an empty intercept with the projection cylinder and no distance calculation. Note also how $\sigma_1(d)$ measured along the normal direction at scale $D_2$ varies with small scale variations in surface orientation which is propagated into a larger local confidence interval.

### *3.1.3 Step 2 : Distance calculation between the two clouds*

Once the normal is defined for the core point $i$, it is used to project $i$ onto each cloud at scale $d$ (called projection scale). This amounts to defining the average positions $i_1$ and $i_2$ of each cloud in the vicinity of $i$ (fig. 3a). This is done by defining a cylinder of radius $d/2$ whose axis goes through $i$ and which is oriented along the normal vector (fig. 3a). A maximum length of the cylinder is imposed to speed up the calculation. The intercept of each cloud with the cylinder defines two subsets of points of size $n_1$ and $n_2$. Projecting each of the subsets on the axis of the cylinder gives two distributions of distances (with an origin on $i$). The mean of the distribution gives the average position of the cloud along the normal direction, $i_1$ and $i_2$, and the two standard deviations give a local estimate of the point cloud roughness $\sigma_1(d)$ and $\sigma_2(d)$ along the normal direction. If outliers are expected in the data (such as vegetation), $i_1$ and $i_2$ can be defined as the median of the distance distribution and the roughness is measured by the inter-quartile range. The local distance between the two clouds $L_{M3C2}(i)$ is then given by the distance between $i_1$ and $i_2$ (fig. 3a).

Fig 3b illustrates a case in which the cloud orientation at scale $d$ is not orthogonal to the normal previously estimated at scale $D$. The 'apparent' roughness $\sigma_1(d)$ will be higher than the 'true' locally detrended roughness. This will yield a larger confidence interval (see section 3.3) consistent with the larger uncertainty associated with a measurement in which the surface orientation is not locally consistent with the normal direction. Note also that if no intercept with the compared cloud is found due to missing data or changes in surface visibility, no calculation occurs (fig. 3b).

### **3.2 Normal scale selection in relation to roughness**

In complex rough surfaces, figure 3b shows that if $D$ is of similar scale (or smaller) than roughness elements, the normal orientation will strongly fluctuate resulting in a potential overestimation of the mean orthogonal distance between the two clouds. Depending on the user application, these fluctuations might be desirable: if one is interested in detecting change of the shape of the meter size boulders on the rockfall example (fig. 1b), a locally small scale $D$ (~ 10-20 cm) would be needed. But in many - arguably most - cases where no corresponding elements can be identified between surveys, we are interested in changes along a normal direction that is not affected by the surface roughness. For the rockfall example, this would mean choosing $D > 10$ m. On the other hand, the scale at which the normal is estimated must be small enough to capture large scale changes in surface orientation (e.g. the transition between river bed and banks, or the change in cliff orientation related to meander curvature in fig. 1a). A key aspect of the point cloud comparison problem in complex topographies with variable roughness is thus to define the optimal scale $D_{opt}$ at which the normal should be estimated.

To our knowledge, this problem has only been approached in the context of smooth slightly curved surfaces affected by a random normally distributed measurement noise (Mitra and Nguyen, 2003;

Lalonde et al., 2005; Bae et al., 2009). In that case an optimum scale $D_{opt}$ can be semi-empirically defined as a function of the noise amplitude, surface curvature and point density. However the semi-empirical models developed in these studies (Mitra and Nguyen, 2003; Lalonde et al., 2005; Bae et al., 2009) need a phase of calibration, may require an estimate of surface curvature (which significantly impacts computation time) and are quite complex. Most importantly, they were not developed in the context of complex natural surfaces for which the roughness might not be distributed along a Gaussian and is characterized by partial sampling owing to significant occlusion.

We have thus implemented a simpler empirical approach: over a range of scales imposed by the user (e.g. 0.5 to 15 m with 0.5 m intervals), we chose the scale at which a plane best fits the 3D surface (i.e. the surface appears the most planar at this scale). For this, we perform a Principal Component Analysis of the neighbours of point $i$ within a sphere of radius $D/2$ and choose the scale $D_{opt}$ at which the third component is the smallest (see (Brodu and Lague, 2012) for details on the calculation). We ensure that a minimum of 10 points is used to compute the normal at $D_{opt}$ otherwise we choose the scale immediately larger that verifies this criteria. It will be demonstrated in section 5.1 that for the rough natural surfaces of the Rangitikei river this mode of selection results in a potential error on distance measurement that is smaller than 2 % of the measured distance, 97 % of the time. The user can also impose a uniform value of $D$ for simplicity and speed. To verify that the normal orientation is unaffected by roughness at the chosen scale $D$, we demonstrate in section 5.1 that $D$ should be at least 20 to 25 times larger than the roughness $\sigma(D)$.

### 3.3 Spatially variable confidence interval

Inspired by recent development in the DoD technique (e.g. Brasington et al., 2000; Lane et al., 2003; Wheaton et al., 2009; Schürch et al., 2011) and the C2M approach (Gosliga et al., 2006), our objective is to define a spatially variable confidence interval associated with each measurement and combining the different sources of uncertainties described in section 2.4. This confidence interval is defined at a prescribed confidence level (95% in the following) and is used to estimate locally the distance measurement accuracy (i.e. 3±5 mm). It is also used to assess whether a statistically significant change is detected or not at the prescribed confidence level. For instance, 3 ± 5 mm is not a statistically significant change at 95 % confidence, while 16 ± 5 mm is. Because the confidence interval boundary corresponds to the minimum detectable change, it is also referred as the Level of Detection at $x$ % ($LOD_{x\%}$[1]) (e.g. Lane et al., 2003; Wheaton et al., 2009). We use interchangeably the terminology *c.i.* (confidence interval at 95%) or $LOD_{95\%}$. In order to construct the $LOD_{95\%}$ indicator we propose two methods. The first one relies on statistical bootstrapping (Efron 1979) and can in principle cope with any type of error distribution. The second one is parametric and based on Gaussian statistics. We describe it below and use it subsequently since it appears to be faster and as accurate on natural scenes. Bootstrapping is detailed in Appendix B.

As the spatial variations of the positioning and the registration errors cannot be modelled easily (see section 2.4), we base the construction of the $LOD_{95\%}$ on the registration error *reg* and the local point cloud roughnesses $\sigma_1(d)$ and $\sigma_2(d)$ measured along the normal direction. The registration error *reg* is hereby assumed isotropic and spatially uniform. $\sigma_1(d)$ and $\sigma_2(d)$ are computed on the two sub-clouds of diameter $d$ and size $n_1$ and $n_2$ that are used to define the average positions $i_1$ and $i_2$ (fig. 3). $\sigma_1(d)$ and $\sigma_2(d)$ depend on the real surface roughness, the correct orientation of the normal with respect to the considered cloud and instrument related noise (e.g., range noise, low incidence angle errors, or mixed point, Soudarissanane et al., 2007, 2011; Hodge, 2010). To estimate the $LOD_{95\%}$, we use a parametric estimate based on Gaussian statistics. The two distributions of distances along the normal direction of estimated mean ($i_1$, $i_2$) and size ($n_1$, $n_2$) are assumed to be independent Gaussian distributions with two potentially different variances estimated by ($\sigma_1(d)^2, \sigma_2(d)^2$). If $n_1$ and $n_2$ are larger than 30, the $LOD_{95\%}$ can be defined for the difference between $i_1$ and $i_2$ by (e.g. Borradaile, 2003):

$$LOD_{95\%}(d) = \pm 1.96 \left( \sqrt{\frac{\sigma_1(d)^2}{n_1} + \frac{\sigma_2(d)^2}{n_2}} + reg \right). \quad (1)$$

---

[1] We use $LOD_{x\%}$ to avoid any confusion with the abbreviation LOD used in the computer graphics community for "Level of Detail".

For a different level of confidence $x$, 1.96 in eq. (1) is replaced by the two-tailed z-statistics at $x\%$. If $n_1$ or $n_2 < 30$, 1.96 in eq. (1) is theoretically replaced by the two-tailed t-statistics with a confidence level of 95 % and a degree of freedom given by (e.g. Borradaile, 2003):

$$DF = \left( \frac{\sigma_1^2}{n_1} + \frac{\sigma_2^2}{n_2} \right)^2 \bigg/ \left( \frac{\sigma_1^4}{n_1^2} \bigg/ (n_1 - 1) + \frac{\sigma_2^4}{n_2^2} \bigg/ (n_2 - 1) \right). \quad (1)$$

However, our tests on natural data (section 5.2) show that eq. (1) provides a good estimate of the confidence interval at 95% as long as $n_1$ and $n_2 > 4$. Hence, we routinely use eq. (1), and consider that below a minimum point number of 4, no reliable confidence interval can be estimated (but a distance measurement is still computed). In section 5.2, we explore how the $LOD_{95\%}$ varies with $d$ on a real case example.

We note that in the case of a perfectly flat surface scanned at normal incidence angle, $\sigma_1(d)$ and $\sigma_2(d)$ are equal to the scanner noise. In this configuration at 50 m, we measured with our scanner (Leica Scanstation 2) $\sigma_1 = 1.41$ mm independent of $d$ (see section 5). Eq. (1) shows that choosing a projection scale $d$ containing 100 pts on each point clouds (e.g., $d \sim 10$ cm for data with 1 cm point spacing), the $LOD_{95\%}$ would be $\pm 0.33$ mm (assuming $reg=0$ mm). This highlights the interest in using the average position of the point clouds to reduce the uncertainty related to scanner measurement noise (Monserrat and Crosetto, 2008; Abellán et al., 2009). While eq. (1) does not directly account for scanner accuracy (i.e. the difference between the actual and measured range distance), the registration error will depend on it (and other elements such as the number and distribution of GCPs or cloud matching surfaces). For instance, if the calibration of an instrument were to deteriorate between two surveys, the registration error would increase and this effect would translate into a higher $LOD_{95\%}$.

### 3.4 Comparison with existing methods on synthetic point clouds
In this section we assess the precision of existing 3D comparison techniques (cloud-to-cloud and cloud-to-mesh) and the M3C2 algorithm. We subsequently apply it to real data and address the effect complex surface roughness in section 5. We use synthetic horizontal surfaces generated numerically with normally distributed noise (standard deviation = 1 mm) and which are vertically shifted by $z_{mean}$. Each cloud contains 100000 points generated with a constant x and y spacing $dx$. We explore two point spacing $dx = 1$ or 10 mm. For each $dx$, several clouds with $z_{mean}$ varying between 1 and 100 mm are generated and compared to a reference point cloud with $z_{mean} = 0$ mm. When dx=10 mm, we also explore the impact of horizontally shifting the second point cloud by 5 mm. We use millimeter units, but these can be replaced by any unit (e.g., cm or m), the results would be identical as there is no absolute scale attached to the algorithm. While this test explores a vertical shift for simplicity, we emphasize that the M3C2 and other 3D algorithms would work in any orientation. We used the open source free software CloudCompare v2.3 (EDF R&D, 2011) to test the various comparison methods (fig. 3):

- C2C: simple nearest neighbour cloud-to-cloud comparison
- C2C_HF: nearest neighbour distance with height function model
- C2M: meshing of the reference cloud and cloud-to-mesh comparison.

The parameters used in the M3C2 algorithm are $D = 50\ dx$ and $d = 10\ dx$. For the cloud-to-mesh comparison, the first phase of normal estimation in CloudCompare was done at a scale equal to $D$. The Poisson reconstruction meshing procedure (Kazhdan et al., 2006) was done at an octree level for which one octree cell has about the same length than the projection normal $d$ in the M3C2 calculation (see (EDF R&D, 2011) for details). This amounts to comparing the two algorithms with the same set of parameters.

Fig. 4a shows the mean change measured by the various algorithms for $dx$=1 mm. The M3C2 and C2M recover on average the vertical displacement very accurately: within 0.003 mm for M3C2 and 0.02 mm for C2M. But the C2M method measurements exhibits a larger variability (1.00 mm vs 0.15 mm), independent of the displacement, which is exactly the standard deviation imposed on the z value. This is expected because the C2M method only averages the position of the reference cloud through the meshing phase (fig. 2) while the M3C2 method averages each cloud. There is a remaining variability because a very small amount of error is introduced by the normal estimate. If we force the normal to be vertical, the standard deviation of the average distance becomes null. We note that when no surface change is imposed, the two methods do not detect a surface change.

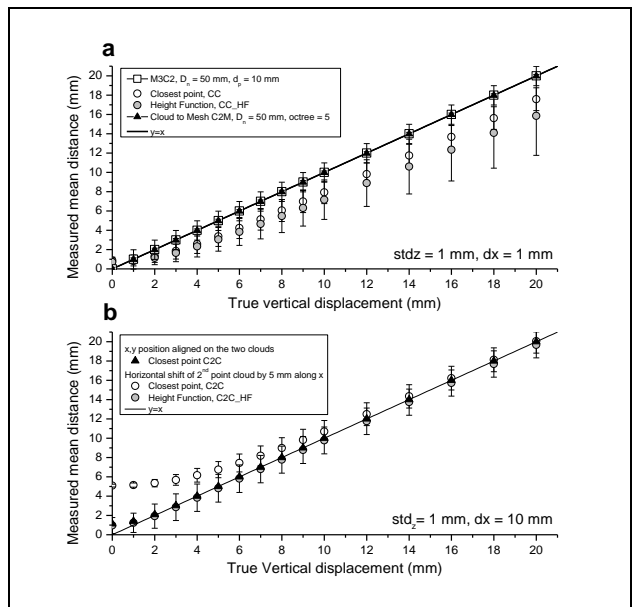

**Figure 4: a :** Measured vs true vertical displacement of an horizontal point cloud with Gaussian noise $std_z$=1 mm and point spacing $dx$ = 1 mm. Errors bars correspond to the standard deviation of the average of 100000 points. Standard deviation for the M3C2 and C2M results are constant and equal to 0.15 mm and 1.00 mm respectively. **b :** closest point calculations with $dx$=10 mm with or without horizontal translation of the points by 5 mm along x. M3C2 and C2M calculations are not shown for clarity as the results are identical than with the case $dx$ = 1 mm.

The cloud-to-cloud measurements based on a closest point calculation (C2C, C2C_HF) fail to capture precisely small surface change when dx = 1mm (fig. 4a) or 10 mm (fig. 4b). The two methods predict a surface change when no change occurs (e.g., for dx = 1 mm, they detect an average change of 0.5 and 0.9 mm depending on the method). The predicted value is a function of the point cloud noise and point spacing: if the point coordinates are not aligned (which is typically the case for surveys at different time), a change up to half the point spacing can be predicted when no change occurs (fig. 4b). For small vertical displacement, the bias can be very significant and is very difficult to predict: for a 4 mm true vertical displacement, $L_{C2C}$=2.56 mm for dx=1mm and 6.16 mm for $dx$ = 10 mm. Note that for $dx$ = 1 mm, C2C increases linearly with the true vertical displacement for displacement larger than 4 mm but is systematically offset by 1.95 mm. This is an effect of the noise of the reference point cloud (fig. 2A). Using a height function to better approximate the point cloud position sometimes offers a significant improvement (fig. 4b, *dx* =10 mm) but can be less precise than a simple closest point comparison if the roughness is of the same order as the point spacing (fig. 4a, *dx* =1 mm).

We conclude that current algorithms of closest point calculations are prone to unpredictable bias for small surface change detection. They remain however extremely useful for fast 3D detection of changes significantly larger than the point cloud spacing and roughness (Girardeau-Montaut et al., 2005). Cloud to mesh and the M3C2 algorithm offer accurate surface change measurement that is independent of point density and surface roughness provided that a large enough $D$ and $d$ are chosen.

## 4. Acquisition of test data for accurate change detection

In this section we present the techniques used to obtain the smallest co-registration errors possible between surveys and the dataset used to test the M3C2 algorithm.

### 4.1 Study site: the Rangitikei river

The Rangitikei river is located on the North Island of New-Zealand and flows over weakly consolidated mudstones that are cohesive enough to sustain cliffs up to 100 m high (fig.1). It transports predominantly cobbles (5-20 cm) and gravels (1-5 cm) with occasional sandy patches and meter size boulders. It is a meandering river with rapid bedrock cliff erosion in the external part of meander bend evidenced by frequent slips and rockfalls. We surveyed a reach of about 500 m long near Mangaweka (Mangarere road bridge). Five surveys were obtained since 2009 with intervals ranging from 2 months to 1 year (02/2009, 02/2010,12/2010,02/2011,12/2011). The primary objective in doing these surveys was to document any kind of surface change (bed sedimentation/aggradation, bank erosion, cliff failure, vegetation growth). We use the registration information of the different surveys to investigate the expected range of registration errors and will use the february 2009 and february 2011 surveys to apply the M3C2 algorithm.

### 4.2 Field setup and data acquisition

The sites were surveyed in low flow conditions using a Leica Scanstation 2 mounted on a survey tripod with dual-axis compensation always activated. Quoted accuracy from the constructor given as one standard deviation at 50 m are 4 mm for range measurement and 60 µrad for angular accuracy. Repeatability (i.e. precision also known as scanner noise) of the measurement at 50 m was measured at 1.4 mm on our scanner (given as one standard deviation), while accuracy was of the order of 0.2 mm at 50 m (obtained by comparing measured change against precisely known millimetric changes). Laser footprint is quoted at 4 mm between 1 and 50 m.

Four to five scanning positions were chosen on elevated spots on the bank or on the river bed (fig. 1). As it is typical with dynamic environments (Schürch et al., 2011), it was not possible to setup the scanner systematically at the same place but we tried to reoccupy the same position to have about the same geometry of occlusion (Girardeau-Montaut et al., 2005). Note that even if performed at low flow conditions, the river is still 50 to 70 m wide and up to 2-3 m deep preventing a direct scanning of the deepest part of the channel (fig. 1). The typical scanning procedure starts with a full 360° coarse scan at low resolution (~ 10 cm point spacing at 50 m) followed by high-resolution scans of the surface of interest in which we aimed at a point spacing of 10 mm horizontally and 5 mm vertically at 50 m from the scanner. As is the case with TLS, point spacing was highly variable on the scene (fig. 1b). The final registered raw cloud consisted of 50 to 60 M points. Because temperature and pressure were always close to the standard calibration of the scanner (20°c, 1013 mbar), no atmospheric correction was used.

## 4.3 Point cloud registration and vegetation removal

We primarily used a target-based registration using Leica blue and white HDS targets mounted on tripods (for intra-survey registration) or anchored on rocks (for local georeferencing). Leica quotes the error on the measurement of target center at 2 mm (1 standard deviation) at 50m. We systematically had a minimum of 3 targets, and up to 5 targets common to two stations and placed within 75 m. Registration was performed using Leica Cyclone 7.2 (Leica, 2011). Targets were placed at various elevations and distributed spatially as evenly as possible in order to obtain a spatially uniform registration error.

On this site, we managed to install 5 targets bolted on rocks installed in the bedrock or on large blocks buried in the inner bank (fig. 1). The limited availability of really fixed emplacements explains the relatively small number of fixed targets and their non-ideal spatial distribution. The bolts are recessed by 4 cm to protect them and precisely machined adaptors are used to position targets on the bolts with an uncertainty less than 0.1 mm (that we neglect). Because we do not have an independent survey of the fixed targets positions, we used a special approach in which all the stations of all the epochs were registered at once, rather than first registering an epoch, and then co-register subsequent epochs using the fixed targets. This method amounts to remeasuring the network of fixed targets 5 times and reducing the standard error of the epoch registration by $\sqrt{5}$. It also prevents any measurement error in the first survey, to systematically propagate on all subsequent epoch registration if there was some erroneous measurement. Subsequent surveys can be registered by using the average position of the fixed targets calculated from the first 5 epochs. This avoids a full re-registration.

Assembling all the data together allows us to look at the statistics of 231 target positions (fig. 5). Without any sort of target preselection, the mean error is 2.44 mm with 95 % of the targets within 5.47 mm. According to the manufacturer specification, we would expect 95 % of the error position to be within 3.92 mm. Analysis of the targets with error larger than 5 mm show that they come from surveys with adverse weather conditions (wind, rain) and temporary targets mounted on pole which might have slightly shifted between scanning position. Removing these targets lead to a mean error of 2.14 mm and 95% percentile of 3.99 mm consistent with the manufacturer specification. Analysis of the registration error on the bolted targets shows that the registration error is on average 2.34 mm with a minimum of 2 mm and a maximum of 2.97 mm. We retain a conservative value of 3 mm as the registration error we use in eq. (1). No systematic change with time was detectable showing that the instrument kept the same accuracy and that fixed bolts remained truly fixed. We kept a local georeferencing configuration with elevation measured above an arbitrary datum. To our knowledge, this is the smallest registration error reported for TLS-based surface change measurements over 100's meter scale.

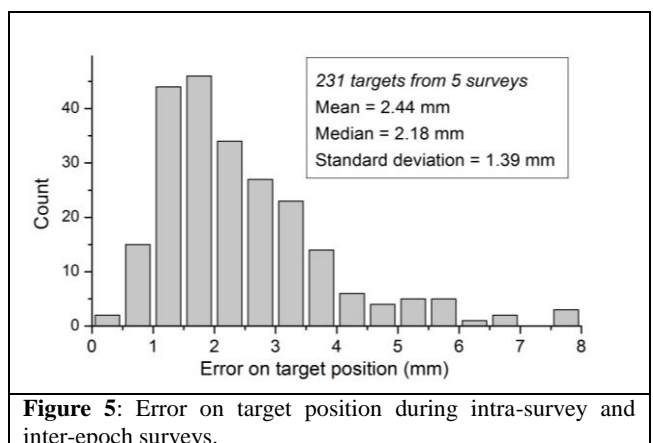

**Figure 5**: Error on target position during intra-survey and inter-epoch surveys.

As explained in Appendix A, we experimented adding point cloud matching constraints with an iterative closest point (ICP) technique (Besl and McKay, 1992) to the target-based registration for the intra-survey registration. But this lead to a systematically worse registration error on the order of 1 cm (as verified by the targets).

Prior to point-cloud comparison, the vegetation is automatically removed using a multiscale dimensionality classification algorithm (Brodu and Lague, 2012) (fig. 1a). The excellent classification rate (on the order of 98%) allows a direct comparison of the surfaces without most of the noise that would otherwise be introduced by vegetation. The February 2009 scene was subsampled to a minimum point spacing of 10 cm to produce our calculation core point file (1.6 M points after vegetation classification). All 3D visualizations are done with Cloudcompare (EDF R&D, 2011) on the point clouds (i.e. no meshing).

We extracted five smaller parts (fig 1b and 1c) to analyze the roughness characteristics of a flat part of the cliff, a cobble bed with 5-20 cm grains, a gravel bed with 1-5 cm grains, a rockfall deposit with meter size boulders and a partially collapsed alluvial bank. These point clouds have a minimum point spacing ranging from 1 to 3 cm. Apart from the flat part of the cliff, the four other surfaces are characterized by significant shadow effects that are typical of ground based surveys of rough surfaces.

## 5. Real case application

This section is divided into 3 parts: we first explore the complex roughness of surfaces found in the Rangitikei river and its impact on the calculation of surface normal as a function of the normal scale $D$. We then explore the validity of the confidence intervals predicted by eq. (1) as a function of surface roughness and projection scale $d$. We calculate the typical level of change detection that can be attained in the Rangitikei river. Finally, we illustrate a typical application of the M3C2 algorithm to the complete scene (fig. 1).

### 5.1 Surface roughness and normal computation

As discussed in section 3.2, surface roughness makes the orientation of surface normal dependent on the scale $D$ at which it is computed (fig 3b). Here we are interested in finding a criteria that would ensure that the scale imposed by the user or chosen by the algorithm (when a range of scales is given) yields a normal orientation that is not affected by smaller scale roughness. We use 5 representative sample clouds of a flat cliff, gravel bed, cobble bed, alluvial bank and rockfall debris (fig. 1c). Since we want to assert the quality of the normal estimation, we tried to choose samples which are the most planar at a large scale, so we know what result is expected in principle, irrespectively of the local roughness. For simplicity, each sample cloud was rotated to be horizontal by fitting a plane (but we stress that the code operates fully in 3D). By working on horizontal surfaces we measure the variability of normal orientation by looking only at the distribution of the vertical component of the normal vector $n.z$: the further it deviates from 1, the larger is the error induced by the roughness on normal estimate. This orientation variability leads to a potential overestimation of the true distance between two clouds. A simple trigonometric analysis predicts that the error is of the order of $E_{norm}(\%) = 100(1-n.z)/n.z$. Here we are interested in finding a criteria that would ensure that the scale $D$ at which we estimate the normal makes $E_{norm}$ negligible (i.e. below 2%). The calculations are made on core points corresponding to a sub-sampled version of each cloud at 10 cm, but using the full point cloud resolution.

Fig. 6a shows how average surface roughness $<\sigma(D)>$ computed for all points of each sample cloud varies with the normal scale $D$. At any given scale, $<\sigma(D)>$ can vary over an order of magnitude between the flat cliff and the rockfall debris (see also, fig. 1b). A simple planar surface with normally distributed noise should have a constant value of $<\sigma(D)>$. Yet, $<\sigma(D)>$ systematically increases with $D$ at different rates depending on the type of surface and the range of scales. There is at least a range of scales in fig. 6a for which the mean roughness can be described by:

$$\langle \sigma(D) \rangle = \sigma(l_0) \left( \frac{D}{l_0} \right)^{\beta}, \qquad (2)$$

where $l_0$ is a reference scale (that we arbitrary choose to be 0.5 m), $\sigma(l_0)$ the roughness at this scale (see fig. 1b) and $\beta$ an exponent (called the Hurst exponent, (Feder, 1988; Renard et al., 2006)). Eq. (2) characterizes self-affine surfaces and a complex roughness organization for a range of scale that is typically of interest for geomorphological applications (~0.1 to 10 m). Table 1 shows that $\beta$ ranges between 0.1 and 0.88, and $\sigma(0.5)$ varies between 5.15 and 71.49 mm. It is behind the scope of this study to present a complete description of the surface roughness characteristics (including any form of anisotropy, (e.g., Butler et al., 2001; Aberle and Nikora, 2006; Renard et al., 2006; Hodge, et al., 2009b). However, recognizing the self-affine nature of these surfaces highlights how scale dependent is any form of measurement made on rough natural surface, including normal and surface change calculation or the estimate of confidence intervals.

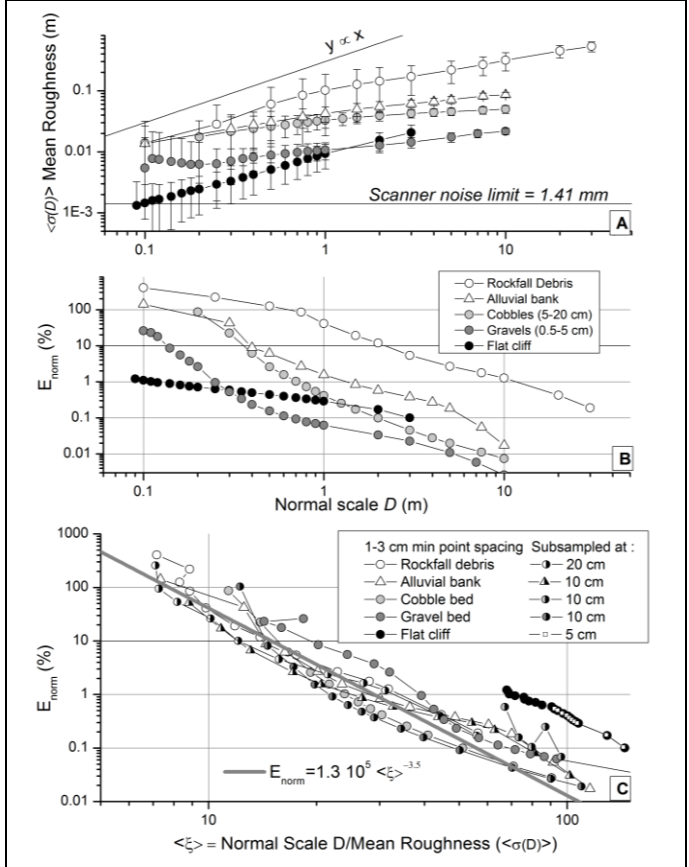

**Figure 6**: **Relationships between scale, roughness and distance overestimation $E_{norm}$ (%) due to deviation of the estimated normal from orthogonality**. 5 different natural surfaces of the Rangitikei river are used (see fig. 1b and text). Data and error bars correspond to the mean and standard deviation of the calculations on core points. Core points numbers : rockfall debris = 19034, alluvial bank = 2304, cobble bed = 34300, gravel bed = 2720, flat cliff = 7188. **A**: Variation of the mean roughness with the normal scale $D$. Note that $D$ is never larger than the minimum horizontal dimension of the point cloud. **B**: Mean distance error due to roughness affecting the estimation of normal orientation as a function of $D$. **C**: Mean distance error as a function of $\xi$, the scale of normal calculation normalized by the roughness at the same scale. Note that the relationship between the error and $\xi$ is not sensitive to the minimum point spacing.

Figure 6b shows that the mean error on distance measurement $E_{norm}$ systematically decreases with the scale of normal computation. This is consistent with the fact that roughness increases less than linearly with $D$ (Fig. 6a). With a threshold error of 2 %, fig. 6b shows that the smallest scale verifying this criteria is 0.25 m for the gravels, 0.70 m for the cobbles, 1.3 m for the alluvial bank and 11.5 m for the rockfall deposits. The flat cliff always fulfils this criteria as long as the scale is larger than 10 cm (given the min point cloud spacing of 2-3 cm for this sample). This highlights the intuitive notion that the scale at which the normal should be estimated depends on the surface roughness (Mitra and Nguyen, 2003; Lalonde et al., 2005; Bae et al., 2009). To factor in this effect,

|  | Range of Scale (m) | Hurst Exponent $\beta$ | Ref roughness, $\sigma$ (0.5m) |
|---|---|---|---|
| Cliff | 0.09-3 | **0.79** ± 0.06 | 5.15 mm |
| Gravel bed | 0.4-10 | **0.31** ± 0.06 | 8.60 mm |
| Cobble bed | 0.2-1 / 1-10 | **0.33** ± 0.09 / **0.17** ± 0.09 | 26.04 / 30.60 mm |
| Alluvial bank | 0.4-10 | **0.31** ± 0.01 | 34.97 mm |
| Rockfall Debris | 0.1-1 / 1-20 | **0.88** ± 0.04 / **0.49** ± 0.07 | 57.30 / 71.49 mm |
| **Table 1**: Parameters of eq. (2) adjusted to the various scaling relationships observed on fig. 6a | | | |

fig. 6c shows the relationship between $E_{norm}$ and a rescaled measure of the normal scale $\xi(i)$:

$$\xi(i) = \frac{D}{\sigma_i(D)}. \quad (2)$$

$\xi(i)$ is the normal scale divided by the roughness measured at the same scale around $i$. This simple rescaling collapses the behaviour of the $E_{norm}$ error into an approximate power-law:

$$E_{norm}(\%) \sim 1.310^5 <\xi>^{-3.5}. \quad (2)$$

$\xi$ thus appears as an indicator of the accuracy of normal orientation. According to eq. (5), choosing $E_{norm} < 2\ \%$ corresponds to $\xi \sim 20\text{-}25$.

The reason for the offset of the cliff data is not clear but does not matter as $E_{norm} < 2\ \%$ for all scales at which a normal scale can be estimated (i.e. the algorithm imposes a minimum of 10 points). Fig. 6c also shows that eq. (2) is not sensitive to a spatial sub-sampling of the data with a 5 to 10 ratio. This has important practical implications as a subsampled version of the point cloud can be used to dramatically speed up the phase of normal calculation in the algorithm compared to using the raw data.

The rescaled normal scale $\xi$ offers a simple way to assess how our algorithm of optimal scale selection based on the most planar scale copes with roughness variations. In fig. 7a, we apply the normal calculation algorithm to a subset of the cliff centered on a large rockfall debris. We used a sub-sampled core point cloud at 10 cm and a range of scale spanning 0.5 m to 15 m with 0.5 m intervals. Fig. 7 shows the resulting normal orientations, optimal normal scale values $D_{opt}$ and $\xi$. As intuitively expected, $D_{opt}$ tend to be small (0.5-2 m) in high curvature regions where the normal tracks changes in surface orientation (e.g. on the overhanging part of the cliff). Large scales values (7-15 m) are selected on the debris and on parts of the cloud where bits of vegetation were not perfectly classified. The flat part of the cliff results in small to intermediate scales. Values of $\xi$ range from 9 to 800, with 97 % of the points characterized by $\xi > 20$ and a corresponding $E_{norm} < 2\ \%$. This shows that our basic approach to the selection of an optimal scale offers a good balance between tracking large scale variations and handling roughness at smaller scale. This mode is however 3 to 4 times longer than with a uniform fixed scale. We also note that using a small lower limit for the range of normals (0.5 m), may result in undesired behaviour: for instance, in the debris zone, the normal for meter size boulders was defined locally at 1 m which yields orientation that are very different than the surrounding parts measured at 15 m; on the rockfall overhanging scar, the calculated normal can be locally parallel to the cliff which might significantly overestimates the distance calculation. A solution to avoid this is to use a higher lower bound, or simply use a uniform large scale (i.e., D= 15 m) that ensures a smooth variation of the normal orientation and prevent any false local incorrect orientation of the normal (which happens in fig 7a for about 0.5 % of the points). To some extent, given the variety of cases found in nature, it is unlikely that a unique solution can be applied for all normal calculations. It is down to the end-user in a specific application context to choose between a variable scale normal estimate or a faster uniform large-scale normal estimate.

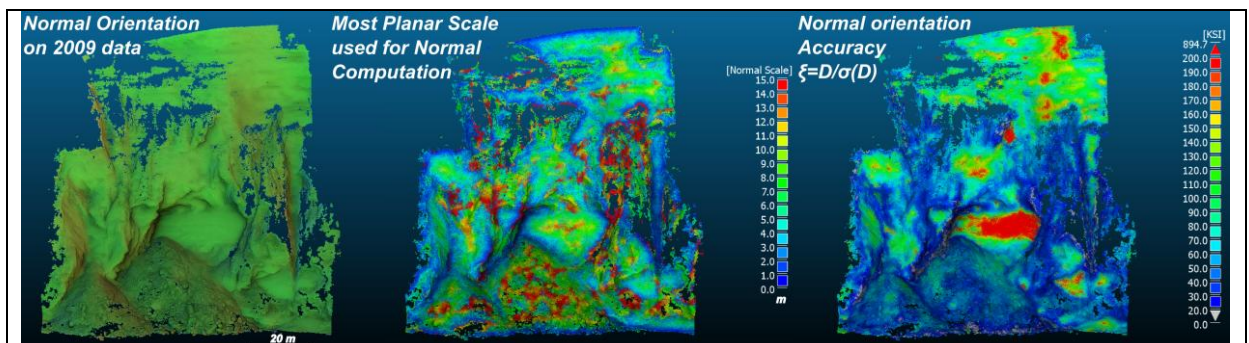

**Figure 7**: **Normal calculation with automatic selection of the most planar scale on the rockfall area** (fig. 1). Normal orientation is defined in a Hue Saturation Value color wheel. $\xi$ is an indicator of normal orientation accuracy given by eq. (4): when $\xi \geq 20$ error on distance calculation between two clouds due to the normal orientation inaccuracy is lower than ~2%. Points in grey correspond to $\xi < 20$.

## 5.2 Confidence intervals for the distance calculation
In the following section, we address two important questions:

- is there an optimal scale of projection $d_{opt}$ with respect to cloud roughness and point density that would minimise the confidence interval while keeping a good spatial resolution ?
- what are the source of uncertainties that dominate the final $LOD_{95\%}$ ?

In general, for natural surfaces, there is no external reference documenting precisely by which amount the surface has changed. In order to assert the quality of the $LOD_{95\%}$ indicator we thus test its ability to correctly detect that a surface has not changed. This is indeed one of the most important aspect of $LOD_{95\%}$ estimates. For this, we use high resolution scans of different rough surfaces and for each of them compare various sub-sampled version of the same data. This way we know that the predicted confidence interval at 95 % should not detect a change 95 % of time. The sensitivity of the $LOD_{95\%}$ to changes in point density between surveys can also be explored. For clarity of the figures we use only 3 of the previously studied surfaces: the flat cliff, the cobble bed and the rockfall debris. Note that the raw point density of these surfaces is 3 times higher for the cobble sample than for the other (650 pt/m² for the cobbles, 191 pt/m² for the cliff and 212 pt/m² for the debris). The subsampled versions of the scenes have similar point densities (5 cm or 4.5 cm gives ~ 62 pts/m², and 10 cm gives ~ 16 pts/m²). In all subsequent calculations, we assess the effects of roughness, scale and point density on the variations of the $LOD_{95\%}$. We impose a fixed normal direction to avoid any effect due to normal misorientation.

### *5.2.1 Projection scale and confidence interval quality*
We investigate how frequently the parametric $LOD_{95\%}$ defined by equations 1 (n>30) and 2 (n<30) detects the absence of change in surfaces that have not changed, but are sampled differently. Fig. 9 shows the percentage of correct rejection of 'significant change' as a function of $d$, for two different sub-sampling configuration. In the first case, we compare the raw data with a subsampled version of it at 10 cm (in that case all the points in cloud 2 exists in cloud 1). In the second case, we produced two subsampled versions: one with a minimum point spacing of 5 cm, and one with a minimum point spacing of 4.5 cm. In that case, around 30 % of the points are identical in both clouds.

Fig. 8 shows that the confidence intervals estimated by the two parametric estimates give similar results except at small values of $d$ (0.1-0.2 m). These values correspond to $n_1$ or $n_2 < 4$. Eq. (1) (which is only strictly valid if $n_1$ or $n_2 > 30$) predicts $LOD_{95\%}$ that are too small. Yet, if $n_1$ and $n_2 \geq 4$, all three predictors of the $LOD_{95\%}$ have similar discriminatory capacity. We thus systematically use a fast parametric estimate with eq. (1) and consider that if $n_1$ or $n_2 < 4$, the uncertainty on surface change is too large and cannot be precisely computed : we consider these measurements as non-statistically significant. Fig. 8 also shows that in many cases the predicted $LOD_{95\%}$ is too large (i.e. conservative) yielding percentages close to 100 %. This means that even if we neglect potential spatial correlation effects between the two surfaces in eq. (1) the $LOD_{95\%}$ is already large enough (Fuller et al., 2003).

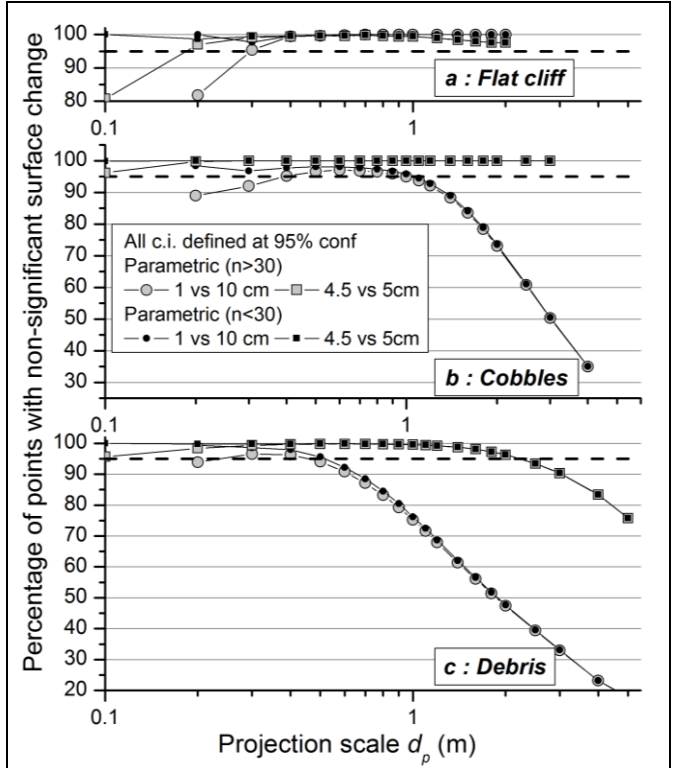

**Figure 8:** Observed percentages of correct significant change rejection for the comparison of identical point clouds with different spatial sub-sampling values. The confidence intervals were defined with a 95% confidence using two different estimates: parametric assuming n>30 (eq. 1) and parametric assuming n<30 (eq. 2). If the predicted confidence was correctly estimated, then the observed percentage of points with no statistically significant change should be 95%.

Most importantly, we observe that increasing $d$ on large roughness surfaces (cobbles and debris) reduces the accuracy of the predicted $LOD_{95\%}$. In particular, when comparing the raw vs 10 cm point clouds, if $d > 0.5$ m, the $LOD_{95\%}$ is too small and predicts a significant change too often. For instance, if $d = 4$ m, about 80 % of the points are incorrectly detected as measuring a significant change. In the debris case, comparing data of similar resolution (4.5 vs 5 cm min subsampling) gives a similar results at scales larger than 2 m. Because eq. (1) supposes that point cloud roughness is normally distributed and that the rough surfaces documented are not (fig. 6), we have also implemented a nonparametric estimate of $LOD_{95\%}$ via bootstrapping (Appendix B). This estimate does not rely on Gaussian statistics. Yet, it yield similar estimate of $LOD_{95\%}$ than the parametric estimate (eq. 1). Hence, we have no explanation for this behaviour nor specific empirical model to correct for it. This effect seems to increase with the surface roughness and the ratio between the point densities of the two clouds. Given that our successive surveys of the Rangitikei have similar point densities, this analysis suggests that the projection scale $d$ must be below 2 m and larger than 0.3 m to stay in the regime were eq. (1) correctly estimates the *c.i.* at 95 %.

### 5.2.2 Scaling of the level of change detection with projection scale

In this section we explore how the $LOD_{95\%}$ predicted by eq. (1) would vary with the projection scale for different surfaces. We compare identical scenes and investigate two point densities (raw vs raw and 4.5 vs 5 cm). Fig. 9 shows that the mean $LOD_{95\%}$ decreases systematically with $d$ for the three surfaces. This can be explained by the dependency of the $LOD_{95\%}$ with $n(d)$ and $\sigma(d)$ in eq. (1). Analysis of the various samples shows that $n$ roughly scales as $d^{1.9}$. Combining this result with eq. (3) and eq. (1) predicts that:

$$LOD_{95\%}(d) \approx LOD_{95\%}(l_0)\left(\frac{d}{l_0}\right)^{\beta-0.95}, \qquad (2)$$

with $l_0$ a reference scale chosen as 0.5 m in our case. $LOD_{95\%}(l_0)$ depends on the roughness at 0.5 m and the point cloud density. Eq. (2) predicts that the $LOD_{95\%}$ will decrease less rapidly with $d$ as $\beta$ increases. According to Table 1, the $LOD_{95\%}$ should decrease roughly as $d^{-0.16}$ for the cliff, $d^{-0.62}$ for the cobbles and $d^{-0.07}$ for the debris. These different sensitivities to $d$ are observed in fig. 9 although no single power-law relationship is observed for the debris. Fig. 9 shows that for the debris and the cliff, there is a very limited decrease in the level of detection by increasing the projection scale. Given that any increase of $d$ will degrade the spatial resolution of the measurement, there little interest to increase *d* for a marginal increase in the capacity to detect a change. This is fundamentally different than for a planar surface only affected by a Gaussian instrumental noise (i.e., $\beta=0$): the $LOD_{95\%}$ then decreases rapidly as the inverse of *d*.

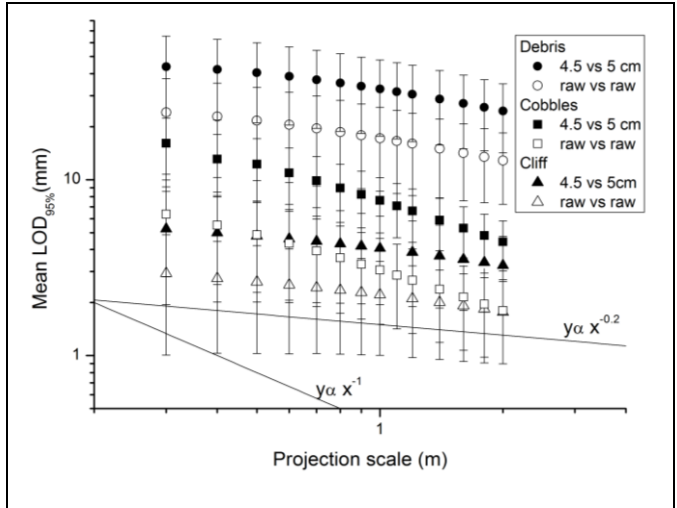

**Figure 9: Parametric mean level of detection at 95% estimated using eq. (1), as a function of the projection scale *d*.** Co-registration error is set to zero. Note that at raw resolution, the debris and cliff have about the same point density (at $d=0.5$, $<n>_{cliff}$ ~ 150 pts, $<n>_{debris}$ ~ 167 pts), but the cobble have a 3 times higher density (at $d=0.5$, $<n>_{cobbles}$ ~ 510 pts). Only scales for which the three $LOD_{95\%}$ estimates are similar and larger than 95 % are shown (see fig. 9).

Figure 9 shows that the point density is an important, and well-known, factor controlling the $LOD_{95\%}$ with a simple dependency as $1/\sqrt{n}$ for a given scale. Increasing the scanning resolution is thus relevant for detecting smaller changes. We note that in the context of rough surfaces, it is likely more important to have a spatially dense measurement (i.e. one scan at very high spatial resolution), rather than a coarser scan with more precise individual measurements (i.e., several identical scans averaged together). The former case offers a better sampling of the surface roughness characteristics.

From the previous analysis and with respect to the questions raised in section 5.2, we conclude that :

- a parametric confidence interval estimate based on Gaussian statistics (eq. 1) can be used on rough surfaces exhibiting self-affine characteristics but on a limited range of projection scales.
- There is a limited improvement on the $LOD_{95\%}$ obtained by increasing *d* but a detrimental effect on the spatial resolution of the computation
- Consequently, the optimum projection scale should be defined such that it verifies $n_1$ and $n_2 \geq 4$ (and preferably $n_1$ and $n_2 \geq 30$ for better statistics) for most of the scene. In the Rangitikei river, we chose $d=0.5$ m.

### 5.2.3 Dominant factors controlling the $LOD_{95\%}$

As discussed in section 5.2.1, the error associated with the normal calculation due to roughness can be neglected as long as $\xi>20$. The contribution of surface roughness and position uncertainty can be estimated from fig. 9 for $d = 0.5$ m. The $LOD_{95\%}$ are of the order of 2.6 mm for the cliff (min = 0.48 mm, max = 10.2 mm), 4.85 mm for the cobble bed (min = 0.9 mm, max = 36 mm) and 21.6 mm for the debris (min = 1.3 mm, max = 260 mm). For the cliff, the min value of the $LOD_{95\%}$ is limited by the range noise of the scanner and the scan density. Otherwise, the surface roughness is always a larger contributor to the final $LOD_{95\%}$ than the instrument uncertainties (with our equipment). Our co-registration standard errors are of the order of 2-3 mm (see section 4.3) which translates into a contribution to a 95% confidence interval of ~ 4-6 mm. Hence, for $d= 0.5$ m, the co-registration error is generally dominant or on par with roughness effects for the cliff and cobble beds at their raw resolution (fig. 10), but is negligible compared to the roughness component for the rockfall debris.

### 5.3 Typical application on the Rangitikei river data

We illustrate the fastest calculation possible which combines the following elements:

- Cliff and inner bank were separated and run in parallel with core point files based on the 2009 data subsampled at 10 cm (1.6 M points).
- Vegetation was removed using the canupo software (Brodu and Lague, 2012). This improves normal estimation quality and reduces computation time as the M3C2 algorithm is only applied on non-vegetated surfaces.
- Normal calculations were done at a fixed scale ($D$=15 m on the cliff, $D$=12 m on the bank) using the core point file. Posterior verification showed that $\xi > 20$ for 98 % of the points.
- Following the analysis in section 5.2, a projection scale of $d$=0.5 cm was used. To speed up the calculation we used a two pass algorithm: a first pass using a maximum length of the projection cylinder of $L$= 1 m for which 86% of the 2009 core points were projected on the 2011 cloud. A second pass with $L$=30 m on the 14 % non-projected core points. After these passes, 8% of the 2009 core points could not be projected on the 2011 data due to missing data. This two pass calculation improves by a factor 4 the calculation time compared to using $L=30\ m$ on all the core points.

All these steps are automatically scripted and the calculation took ~ 1 hour on a mobile core-i7 CPU. Figure 10 shows the inputs and outputs of the algorithm. Note that the two surveys had significant difference in surface visibility: during the 2011 survey, the water level was 20 cm lower which led to a larger bed exposure, and the cliff was drier which led to a higher number of laser returns at higher elevations. But the vegetation pattern was also different which led to area exposed in 2009 that were no more visible in 2011. These differences in surface visibility are however well accounted for by the algorithm which do not compute a difference if no projection is found in the 2011 data along the normal direction computed in 2009. Hence, there is no need for manual trimming of the data.

The vertical component of the estimated normal vector smoothly tracks large scale changes in surface orientation up to overhanging parts (dark blue in fig. 10c). Because in this example we used a fixed value of $D$, the small variations in surface orientation observed in fig. 7a are smoothed out. However, the normal correctly rotates from vertical on the river bed to near horizontal on the channel banks. The map of $LOD_{95\%}$ highlights the impact of surface roughness with low values on cliffs and high values on the debris. It also illustrates the effect of point density with sharp transitions in area of scan density changes. The median $LOD_{95\%}$ is 1.87 cm and the median values of $n_{2009}$ and $n_{2011}$ at $d$=0.5 m are 97 and 61 on the cliff, and 84 and 29 for the bank. The map of significant changes (i.e. $n_{2009}$ and $n_{2011} \geq 4$ and $L_{M3C2} > LOD_{95\%}$) shows that most of the surface changed between 2009 and 2011 (by at least 5.9 mm, the lowest $LOD_{95\%}$). The non-significant changes (24 %) corresponds to areas with very low point density (i.e. high cliff zones), zones in which vegetation is not fully removed and yields locally high values of roughness, truly stable zones or areas were the combination of erosion and sedimentation resulted in no net change. We checked that the blocks with fixed reference targets did not show up as significantly changing.

Finally, the map of surface change shows a very large range of detected absolute changes from 6 mm to 6.96 m. We use a logarithmic color scale to account for the 3 orders of magnitude of detected change. The largest recorded change is - 6.96 m of cliff removed in the area of rock failure and an extension up to + 4.51 m of the corresponding rockfall deposits. This large perturbation triggered erosion of the opposite alluvial bank by up to - 2.7 m and an erosion of the bed by - 0.6 m. Several gullies on the cliff deposited up to +2.5 m of sediment at the cliff foot. We document subtle patterns of bed aggradation of up to +0.93 m of gravels and +0.47 m of cobbles. Several smaller rockfalls up to 1 m deep can be recognized in the cliff. We also document a non-uniform background cliff erosion of -5 to -10 cm over 2 years. While relatively large, this rate is consistent with the weak nature of the bedrock material, with the large volume of sediment deposited on the gullies fan and observations during surveys of very frequent small rocks falling down the cliff. We note that zones of non-significant surface changes are intertwined with the zone of background cliff erosion. This show that this background cliff erosion is not an artifact of incorrect registration, but a true surface change. We also document a 3.5 m high band of localized erosion (- 0.3 to - 0.15 m) at the base of the cliff in the meander bend corresponding to the location of higher flow velocity.

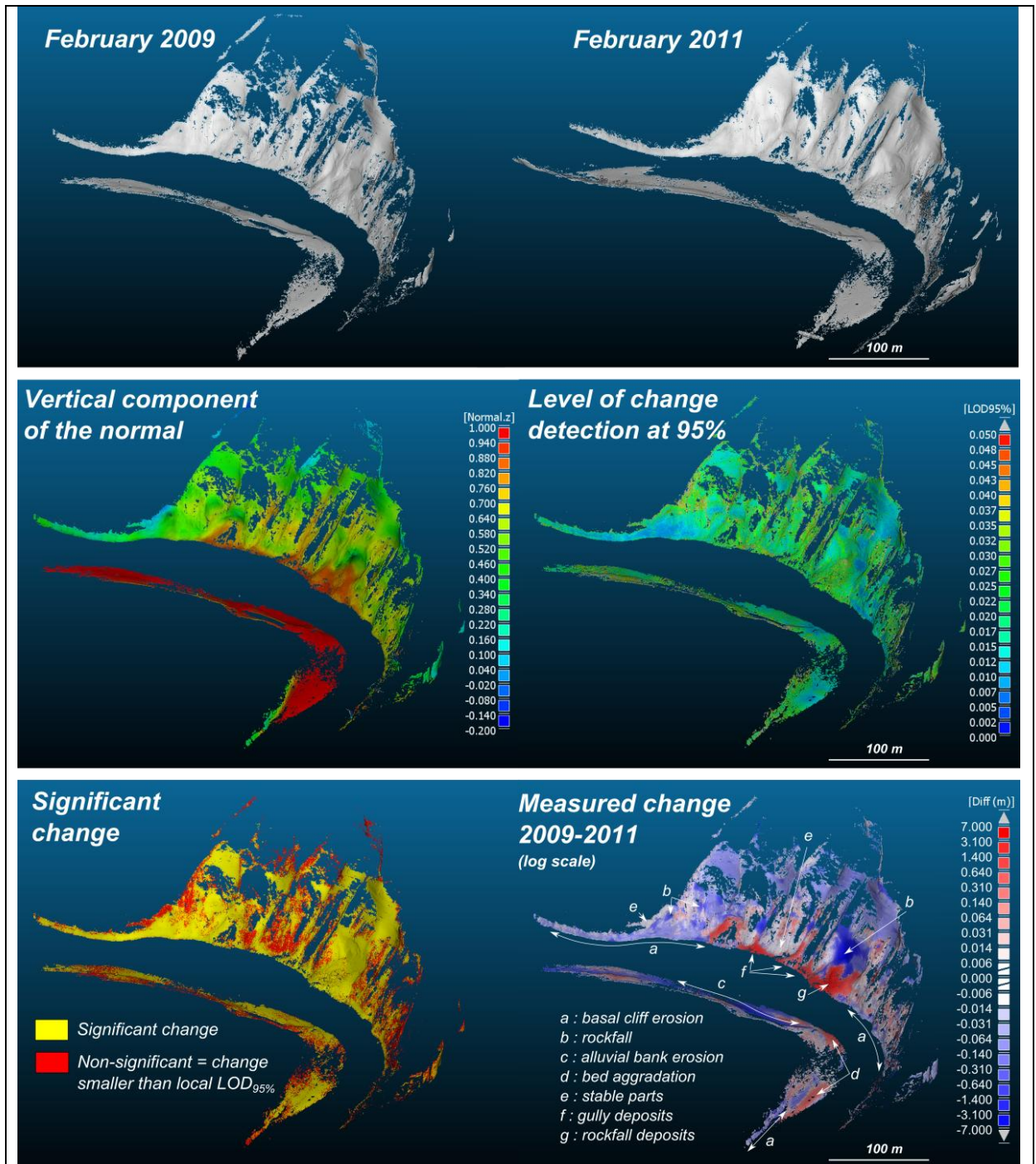

**Figure 10**: **Application of the M3C2 algorithm to the Rangitikei canyon data**. Raw input data are the February 2009 and 2011 surveys shown here without vegetation for clarity. The normal orientation are computed from the 2009 data at a fixed scale (D=15 m for the cliff, and 12 m for the inner bank). The $LOD_{95\%}$ is given by eq. (1) and includes a component of 5.88 mm due to registration uncertainty (1.96 x 3 mm). A change is considered statistically significant when the $LOD_{95\%}$ is smaller than the measured change and that a minimum of 4 points were used to compute the average position of each cloud. Surface change is presented in logarithmic scale with a minimum value of 6 mm corresponding to the minimum detectable change.

# 6 Discussion

## 6.1 Comparison with existing techniques

Compared to existing techniques, the M3C2 algorithm introduces several new elements that both simplify the comparison of point clouds of natural environments in 3D, but also open up the possibility to detect very small surface changes and assess whether they are statistically significant. Key advantages compared to existing techniques such as the Difference of Dem (DOD), cloud-to-cloud (C2C) or cloud-to-mesh (C2M) are:

### 6.1.1 The M3C2 method do not require gridding or meshing of the point cloud.

In the absence of data in the compared cloud (due to changes in occlusion, survey extent or vegetation), the DoD and the C2M method would interpolate the data. The C2C method would generate spurious results (Girardeau-Montaut et al., 2005). The M3C2 algorithm simply does not compute a difference in the absence of an intersection with the compared cloud along the normal direction. The method is thus extremely robust to missing data and greatly simplifies the treatment of point cloud as no manual trimming is needed prior to difference computation. If vegetation is present, automated classification methods now exist (Brodu and Lague, 2012) to remove it efficiently. As the latest generation of point cloud visualization softwares (e.g. Cloudcompare, EDF R&D, 2011) do not require surface mesh for visualisation (fig. 1, 7 and 11 are all produced without mesh), there is no need to systematically mesh the high density clouds generated from TLS or advanced photogrammetric techniques (e.g. Structure From Motion).

### 6.1.2 The M3C2 method operates in 3D.

A lot of the philosophy behind 2D DoD calculations has been used to construct the M3C2 algorithm, in particular the emphasis on spatially variable confidence intervals (Brasington et al., 2000; Fuller et al., 2003; Lane et al., 2003; Milan et al., 2007; Wheaton et al., 2009; Schürch et al., 2011). However, since it fully works in 3D, the M3C2 method alleviates a key limitation of the DoD technique. This makes M3C2 more appropriate for TLS and photogrammetric data in complex 3D environments. Note that the M3C2 method can be directly used as a substitute of the DoD technique for 2D surfaces by using a grid of core points (easily generated with CloudCompare), imposing a vertical normal (which makes calculation extremely fast) and a projection scale $d$ slightly larger than the grid spacing (to approximate the averaging over one square pixel). The resulting calculation is a grid of points with several scalars attached to it (vertical difference, $LOD_{95\%}$, significance of change, roughness....) that can all be turned into raster grids. An interpolation may be needed after the calculation if some core points could not be projected due to missing data, but it is safer to do it at this stage, rather than on the raw data.

### 6.1.3 The M3C2 method is accurate and estimates its local accuracy

Synthetic tests show that the method can accurately recover the mean surface change of noisy surfaces by using an estimate (mean or median) of the local average position of each cloud. The method is thus robust to changes in point density and point cloud noise, which is not the case for the C2C method (Girardeau-Montaut et al., 2005). C2M or mesh-to-mesh techniques can be as accurate as the M3C2 method, but none to our knowledge offers the possibility to locally estimate a confidence interval. The $LOD_{95\%}$ can be used to test for the statistical significance of a measured change given the co-registration error, surface roughness and point cloud density. An indirect, but desirable, consequence is that surfaces with remaining vegetation points will have a high $LOD_{95\%}$ compared to surrounding points. In the case of incorrectly aligned scans obtained from two different viewpoints of the same survey, the resulting point cloud is characterized by an artificial roughness (e.g. Schürch et al., 2011) that can be significantly larger than the true surface roughness (or instrumental noise). Because this roughness is locally estimated and used in the $LOD_{95\%}$ estimate, this incorrect scan alignment is implicitly factored in the comparison process. Although we did not present the results, a robust estimate of the average point cloud position can be obtained by using the median rather than the mean value. This helps removing the impact of outliers such as bits of vegetation that were not fully removed during classification. Estimating a confidence interval in that case requires a slower bootstrap estimate (Appendix B).

### 6.2 Current limits and margins for improvement

Simplifying and improving the accuracy of 3D point cloud comparison comes at the cost of larger calculation times compared to existing techniques (especially the C2C method). The use of core points significantly speeds up the calculation and produces results (fig. 10) that have a resolution more desirable for interpretation than raw data with very pronounced variations in point density. In our test data, given that we use a projection scale of 50 cm (i.e. a radius of 25 cm), there would be little interest in using a core point spacing smaller than 10 cm (which already represents some degree of oversampling).

Our elementary method of normal estimation could be improved by using a curved surface rather than a plane to better track high frequency changes in normal orientation (e.g. Mitra and Nguyen, 2003; Lalonde et al., 2005; Li et al., 2010). However, the estimation of normal orientation on self-affine surfaces is a challenge in itself. The choice of the scale of normal calculation is also depends on the specific user application (e.g., individual blocks vs whole rockfall deposit). In these cases, the metrics ξ offers a posterior assessment of the accuracy of normal orientation for the user-defined scale D (or range of scales). We also note that our choice of external points for orienting the normal might fail in very complex 3D surfaces, and advanced methods for checking the normal orientation consistency might be needed.

The parametric estimate of the $LOD_{95\%}$ (eq. 1) is a first attempt to capture the effects of surface roughness, position uncertainty and registration error. Yet, the position uncertainty will vary as a function of distance from the scanner, incidence angle and surface reflectance (e.g. Soudarissanane et al., 2009, 2011). Similarly the registration will not generally be isotropic and spatially uniform. An improvement of eq. (1) would be to factor in these spatial variations. Another issue raised in our analysis is the need for statistics dedicated to the comparison of self-affine surfaces. Such statistics may explain why a simple Gaussian estimate yields reasonable $LOD_{95\%}$ at small scale but fails at large scales on rough surfaces.

### 6.3 Additional applications of the M3C2 method

The M3C2 algorithm can be used for the simplification of point clouds by local averaging (using the mean) or outlier removals (median). For this a core point file with any desired point spacing and geometry can be constructed and projected on the raw point cloud by comparing the raw data with itself. The projection scale *d* will define the averaging scale. Several scalars of interest can be attached to the core point such as the slope angle (vertical component of the normal vector) (fig. 10), the surface roughness or the point density (fig. 1b). The M3C2 algorithm can also be used to compute normal vectors at the most planar scale on any point cloud.

## 7. Conclusion

This paper has introduced a new method called Multiscale Model to Model Cloud Comparison (M3C2) for direct point cloud comparison in 3D. The algorithm has been designed for accurate orthogonal distance measurement between two point clouds. It differs from techniques based on the matching of homologous parts yielding a displacement field. It significantly improves on existing distance measurement solutions by being 3D, avoiding meshing or gridding steps and being robust to missing data or changes in point density between surveys. Roughness is a key element that is measured and used at every step of the calculation. The normal can be automatically defined at a scale consistent with the local roughness. Local averaging around each point further reduces the influence of the surface roughness. Finally, M3C2 also uses a local measure of the cloud roughness and point density to estimate a parametric confidence interval, and test for the statistical significance of measured changes. To our knowledge, no algorithm has ever combined such a complete array of features to operate in 3D.

The M3C2 algorithm was applied to a 500 m survey of a complex 3D meandering river canyon characterized by large changes in surface roughness and different patterns of occlusion. We showed that the detrended roughness $\sigma(D)$ of various surfaces exhibits self-affine behaviour. This important observation has profound consequences for the choice of the two scales that the user must define to use the M3C2 algorithm :

- For the normal scale $D$, we found that the orientation uncertainty decreases rapidly with $\xi(D)=D/\sigma(D)$ (eq. 5). A rule of thumb is that the uncertainty becomes negligible in the comparison process when $D$ is 20-25 times the roughness estimated at scale $D$. Given that the relationship between $\sigma(D)$ and $D$ depends on the type of surface (fig. 6), there is no way to predict *a priori* the exact scale $D$ above which $\xi(D)>25$. In the case of variable surface roughness the user can thus define a range of scales: the optimal normal scale is automatically chosen as the most planar scale. In our study case, this method yields optimal scales that can vary over nearly two orders of magnitude (0.5 - 15 meters) depending on the surface roughness and geometry. For 97 % of them, the optimal scale verifies $\xi(D)>25$.

- For the projection scale $d$ over which each cloud is averaged, we demonstrated that the self-affine nature of the surface limits the reduction in the level of change detection $LOD_{95\%}$ when $d$ increases. We also showed that for large roughness surfaces (e.g. rockfall debris), the predicted $LOD_{95\%}$ is too small when $d$ is greater than 1-2 m and incorrectly predicts a statistically significant change, when none occurred. These effect could be due to Gaussian statistics being inappropriate for self-affine surfaces. Statistical bootstrapping, supposedly more adequate for these situations, does not provide a better estimate (Appendix B). Considering these effects, $d$ should be chosen to be large enough to average a minimum of 20 pts (e.g., $d \sim$ 15 cm for a point density of 1 pt/cm²) but small enough to avoid a degradation of the measurement resolution by spatial averaging. The exact value of $d$ will depend on the user application, point density and surface complexity.

Registration error based on fixed targets was consistently within 2-3 mm (at 1 std) over 5 surveys covering 3 years. This is roughly 1 order of magnitude lower than previously reported typical registration errors in similar environments. This results from the use of a precise and accurate TLS (Leica Scanstation 2) and GCPs rather than cloud matching techniques for registration (Appendix A). Applying the M3C2 in this context showed that the $LOD_{95\%}$ ranges from 6 mm to ~ 6 cm, of which 6 mm represents the registration error component. The lowest $LOD_{95\%}$ corresponds to flat cliff area where the averaging effect reduces the uncertainty due to instrumental noise and surface roughness to a sub-mm component. The map of distance measurement shows that the M3C2 can efficiently detect changes covering 3 order of magnitudes (6 mm - 7 m) along orientations that track large scale surface variations. It also highlights areas in 3D where the measured change is not significant.

An implementation of the M3C2 algorithm is available as an open source/free software on the main author's website[2] and can be applied to any source of point cloud data (TLS, photogrammetry, GPS, bathymetric data...) and geometry. It can be used in place of the DoD technique on 2D surfaces to avoid a complex and risky gridding phase. Scripting allows for automatic processing of large batch of surveys without supervision. Used in combination with powerful classification softwares (Brodu and Lague, 2012) and visualisation/processing tools (EDF R&D, 2011), it can help unlocking the scientific potential of very large point cloud datasets.

### Appendix A: ICP registration

Tryouts of raw ICP registration only on the whole scene led to significantly poorer registration than with targets. For instance, using targets as reference points (but not in the registration process) lead to position error systematically larger than 1 cm. Choosing manually only densely scanned 3D elements on the scene such as large blocks, overhanging parts of the cliff slightly improved position error on targets, but never to the point that the registration error was better than 5 mm. This is not unexpected as the standard ICP works well for smooth overlapping 3D surfaces scanned at very high resolution (i.e. the corner of a room), but can have a large remaining uncertainty on noisy 2D surfaces (Salvi et al., 2007; Bae and Lichti, 2008). Given that a standard ICP (Besl and McKay, 1992) is based on a closest point measurement, it potentially suffers from the same drawbacks presented in section 3.4 and fig. 4. From a practical point of view, we found that ICP registration is also an overall longer process than target-based registration: while targets have to be moved in the field (but this can be done during scanning), the registration is instantaneous (and can be checked in the field), while ICP registration requires to clean and highlight manually parts of the scene that respect ideal conditions. Moreover, ICP registration cannot be used for high precision surface change measurement in contexts where most of the elements of the scene might actually change (see section 5). We conclude that ICP registration will not yield high-precision registation (i.e., < 1 cm) in natural environment, and that it is best (when possible) to use a local network of fixed targets evenly spread out in the scene on objects that are known to be fixed.

### Appendix B: Bootstrap confidence intervals

Eq. (1) requires several assumptions to be valid. When using the median instead of the mean for projecting along the surface normal, it is even more difficult to define a confidence interval analytically. For these reasons, we have implemented an optional bootstrap estimate of the confidence interval (Efron, 1979). This technique amounts to generating a large number of random resamples of the data, drawn with replacement, in order to estimate the statistic of interest (here the value of the distance and its standard deviation) from these resamples. Under mild conditions the bootstrap value converges to the correct one given a large enough number of resamples. We use the percentile bootstrap estimate at the $x$% confidence level given by the product of the standard deviation of the bootstrapped distance and the z-statistics at $x$% (i.e. 1.96 for $x$=95%). This method is expected to be more precise than the parametric one as it does not rely on any specific assumption about the underlying statistics of the data.

Our test showed that the bootstrap and parametric $LOD_{95\%}$ predicted by eq. (1) when $n>4$ can differ for any given point by 20 % for the flat cliff and cobble, and up to 25 % for the debris. But they are identical when averaged over all the cloud points. Given that the bootstrap approach significantly increases the computation time, we have not found a practical advantage to using it. If a significant proportion of outliers is expected, the median should be used instead of the mean as a way to approximate the central position of each cloud. In that case the parametric LOD is not available and bootstrapping is a way to still get an estimate of the confidence interval.

---

[2] http://www.geosciences.univ-rennes1.fr/spip.php?rubrique95

## Acknowledgments:

Daniel Girardeau-Montaut is greatly acknowledged for his ongoing development of the open source/free software CloudCompare, and his inspirational work on point cloud comparison. The authors are indebted to Stephane Bonnet and Nicolas Le Dantec for field assistance. Jean-Jacques Kermarrec and Pascal Rolland provided technical assistance in Rennes (Fr). Tim Davies, Vanessa Tappenden and Cathy Higgins are acknowledged for their help at the University of Canterbury (N-Z). Philippe Davy provided insights into the statistics of self-affine surfaces. This research was supported by a Marie Curie International Outgoing Fellowship within the 7th European Community Framework Programme (PIOF-GA-2009-254371) to D. Lague. Additional funding was provided by the CNRS/INSU (projet blanc, projet EC2CO) and the University of Rennes through the CPER 2008-2013. We thank the editor and three anonymous reviewers for their careful analysis of this work.